\documentclass[twocolumn,amsmath,amssymb,nofootinbib]{revtex4-2}

\usepackage{subfiles}
\usepackage{graphicx}
\usepackage{dcolumn}
\usepackage{bm}
\usepackage{hyperref}
\usepackage{verbatim}
\usepackage{siunitx}
\usepackage{upgreek}
\usepackage{multirow}
\usepackage{subcaption}
\usepackage{caption}
\usepackage{ragged2e}
\graphicspath{{\subfix{Figures/}}}

\begin{document}

\title{\textbf{A Collimation System Baseline Design for the Electron Storage Ring at the Electron-Ion Collider} 
}

\author{Andrii Natochii}            
 \email{Contact author: natochii@bnl.gov}
\author{Elke-Caroline Aschenauer}   
\author{Karim Hamdi}                
\author{Charles Hetzel}             
\author{Eric Link}                  
\author{Daniel Marx}                
\author{Christoph Montag}           
\author{Steven Tepikian}            
\affiliation{Brookhaven National Laboratory, Upton, New York 11973, USA}

\author{Yunhai Cai}                 
\author{Yuri Nosochkov}             
\affiliation{SLAC National Accelerator Laboratory, Menlo Park, California 94025, USA}

\date{\today}

\begin{abstract}
We present the baseline design of the electron ring collimation system for the Electron-Ion Collider (EIC) at Brookhaven National Laboratory (BNL). The system addresses beam losses in a high-current electron storage ring with superconducting (SC) final-focus magnets and sensitive detectors, where uncontrolled losses can generate heat loads, radiation, and detector backgrounds and damage. The proposed collimation insertion localizes halo particle losses through reducing interaction region beam losses from beam-gas and Touschek scattering by several orders of magnitude while keeping detector backgrounds and cryostat heat loads within acceptable limits. Multi-turn particle tracking simulations show that the collimators do not significantly impact machine acceptance or beam lifetime, and their positions and apertures can be re-optimized for future lattice configurations. Ongoing work includes incorporating crab cavities and solenoid fields into simulations, refining vacuum conditions, and optimizing collimator geometry and materials. This design establishes a robust baseline for the EIC electron ring collimation system and supports continued lattice optimization for machine operations.
\end{abstract}

\maketitle

\section{\label{sec:Introduction}Introduction}
The Electron-Ion Collider~\cite{osti_1765663}, which will be built at Brookhaven National Laboratory (BNL, USA) over the next decade, is designed to collide polarized electrons with polarized protons and ions over a wide range of center-of-mass energies and at high luminosity, enabling studies of nucleon and nuclear structure with unprecedented precision. To achieve luminosities up to \SI{1e34}{cm^{-2}.s^{-1}}, the machine employs a crab-crossing scheme that restores the effective head-on overlap of the bunches, enabling collisions between relatively long ($\sim\SI{7}{cm}$) hadron bunches and short ($<\SI{1}{cm}$) electron bunches.

High-current storage rings and colliders, particularly those incorporating superconducting (SC) magnets and sensitive detectors, face significant operational challenges from uncontrolled beam losses, radiation deposition, heat loads, and detector backgrounds. A robust collimation system is therefore essential to intercept halo particles and beam fragments, to protect SC magnets and cryogenic systems, to reduce activation of accelerator components, to preserve vacuum quality, and to limit backgrounds in the experimental detectors.

For the electron beam circulating in the EIC, the requirements on beam collimation are especially stringent. While SuperKEKB~\cite{10.1093/ptep/pts083} at KEK (Japan) represents the closest operational analogue in terms of high beam current, small emittance, and demanding detector background constraints~\cite{NATOCHII2023168550}, its collimation strategy relies on an extensive system comprising more than ten collimators per ring for a single beam energy, including multiple collimators located within approximately \SI{100}{m} of the interaction point~\cite{PhysRevAccelBeams.23.053501,PhysRevAccelBeams.27.081001}. This distributed layout enables effective upstream halo cleaning before particles reach the interaction region aperture bottlenecks, thereby substantially reducing detector backgrounds. In contrast, the EIC electron ring must accommodate additional constraints near the interaction point arising from polarization preservation requirements, crab-crossing instrumentation, and the shared, space-limited interaction region with the hadron ring. Moreover, the presence of SC elements in the hadron storage ring and common SC cryostat in the interaction region impose stricter tolerances on beam losses. As a result, the EIC collimation concept cannot rely on extensive near-IP collimation and instead requires a carefully engineered system that is coherently embedded into the lattice, optics, and beam dynamics design of the ring, with a stronger emphasis on global loss control and upstream halo cleaning.

Similar challenges in controlling beam-induced losses and protecting sensitive components have been addressed in other accelerator facilities, notably in the multi-stage collimation system of the Large Hadron Collider (LHC, CERN)~\cite{PhysRevSTAB.17.081004,Redaelli:2016collimation} and in recent electron-positron Future Circular Collider (FCC-ee, CERN) collimation design studies~\cite{Broggi:2026collimation}, where machine protection and detector background mitigation also play a central role in shaping the collimation strategy.

This paper presents the baseline design of the collimation system for the EIC electron ring. The design framework includes the lattice configuration and optimized optics across multiple energies; dedicated machine acceptance studies; collimator placement and specification; and comprehensive multi-turn particle-tracking simulations. We report the resulting beam loss distributions and beam lifetime estimates. Finally, we outline areas for further refinement and optimization.

The remainder of this paper is organized as follows. Section~\ref{sec:Introduction} provides an overview of the collider and detector designs and summarizes the major electron beam loss mechanisms and their associated backgrounds. Section~\ref{sec:CollimationSystemDesign} introduces the electron ring collimation system, describing its layout in the tunnel and lattice together with the relevant beam-optics parameters. Section~\ref{sec:MultiTurnParticleTracking} presents the multi-turn particle-tracking framework, assumptions, and input parameters, and also discusses the ring vacuum system in detail. Section~\ref{sec:TrackingSimulationResults} reviews the simulation results, including collimator-aperture scans, their impact on machine acceptance, and resulting beam loss patterns and lifetimes. Section~\ref{sec:FurtherStudiesOutlook} outlines future studies related to beam collimation, such as refining vacuum parameters, incorporating machine error models, and establishing beam loss limits. Conclusions and a summary of the implications of this work are presented in Section~\ref{sec:Conclusions}.

\subsection{Collider Design}

The EIC will reuse the existing \SI{3.8}{km} Relativistic Heavy Ion Collider (RHIC, BNL) tunnel. One of the RHIC rings will be modified to serve as the Hadron Storage Ring (HSR), operating at 41 and 100 to \SI{275}{GeV}. An Electron Storage Ring (ESR) will be installed in the same tunnel to store electrons at 5, 10, and \SI{18}{GeV}. Together, the two rings will provide a center-of-mass energy range of 29 to \SI{141}{GeV} and support instantaneous luminosities up to \SI{1e34}{cm^{-2}.s^{-1}} with highly polarized electron beams and polarized proton and light-ion beams, for which average polarizations of approximately \SI{70}{\%} are expected. The machine will accommodate collisions between electrons and a broad spectrum of ion species, from protons to heavy nuclei.

The EIC will employ the following injection scenario. The hadron beam is first filled and accelerated to its top energy, after which it is stored for a physics run of approximately 10~hours, during which its intensity gradually decreases. Once the hadron beam is at store energy, the electron beam is injected and operated using a continuous ``swap-out'' injection mode, in which individual electron bunches are periodically replaced at a maximum rate of \SI{1}{Hz}. This approach maintains a stable electron current and high polarization throughout the store.

\begin{figure}[htbp]
\centering
\includegraphics[width=\linewidth]{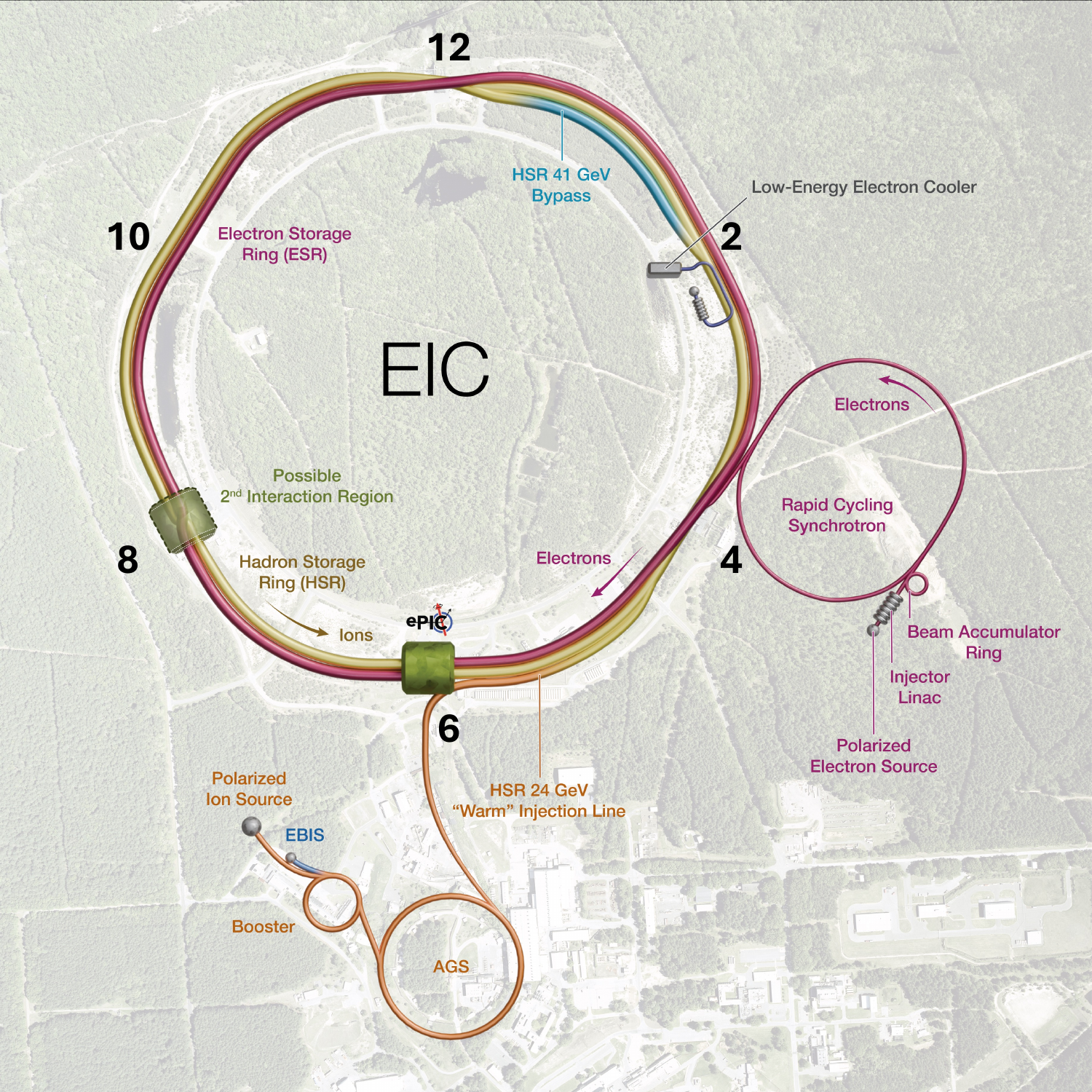}
\caption{\label{fig:EIC_schematic_drawing}Schematic drawing of the EIC. The numbers from 2 through 12 indicate six insertion regions around the rings.}

\end{figure}

The current EIC design includes a single interaction region (IR) and one detector, with a possible future upgrade path to include a second IR (Fig.~\ref{fig:EIC_schematic_drawing}). The key machine parameters relevant to this study are summarized in Table~\ref{tab:MachineParameters}.

\begin{table*}[htbp]
\centering
    \caption{\label{tab:MachineParameters}
    EIC beam parameters relevant to the studies presented in this work, adapted from Table~3.3 of Ref.~\cite{osti_1765663}. The quantities $\beta^{*}$ and $\Delta p/p$ denote the betatron function at the interaction point and the fractional momentum spread, respectively.}
    \begin{tabular}{l|cc|cc|cc}
    \hline\hline
    Species & proton & electron & proton & electron & proton & electron \\
    Energy [GeV] & 275 & 18 & 275 & 10 & 100 & 5 \\
    \hline
    Bunch intensity [$10^{10}$] & 19.1 & 6.2 & 6.9 & 17.2 & 4.8 & 17.2 \\
    No. of bunches & \multicolumn{2}{c|}{290} & \multicolumn{2}{c|}{1160} & \multicolumn{2}{c}{1160} \\
    Beam current [A] & 0.69 & 0.227 & 1 & 2.5 & 1 & 2.5 \\
    RMS emittance, H/V [nm] & 18/1.6 & 24/2 & 11.3/1 & 20/1.3 & 26/2.3 & 20/1.8 \\
    RMS bunch length [cm] & 6 & 0.9 & 6 & 0.7 & 7 & 0.7 \\
    RMS $\Delta p/p$ [$10^{-4}$] & 6.8 & 10.9 & 6.8 & 5.8 & 9.7 & 6.8 \\
    $\beta^{*}$, H/V [cm] & 80/7.1 & 59/5.7 & 80/7.2 & 45/5.6 & 61/5.5 & 78/7.1 \\
    Luminosity [$\mathrm{10^{33}~cm^{-2}s^{-1}}$] & \multicolumn{2}{c|}{1.54} & \multicolumn{2}{c|}{10} & \multicolumn{2}{c}{3.68} \\
    \hline\hline
    \end{tabular}
\end{table*}

\subsection{Detector Design}

\begin{figure}[htbp]
\centering
\includegraphics[width=\linewidth]{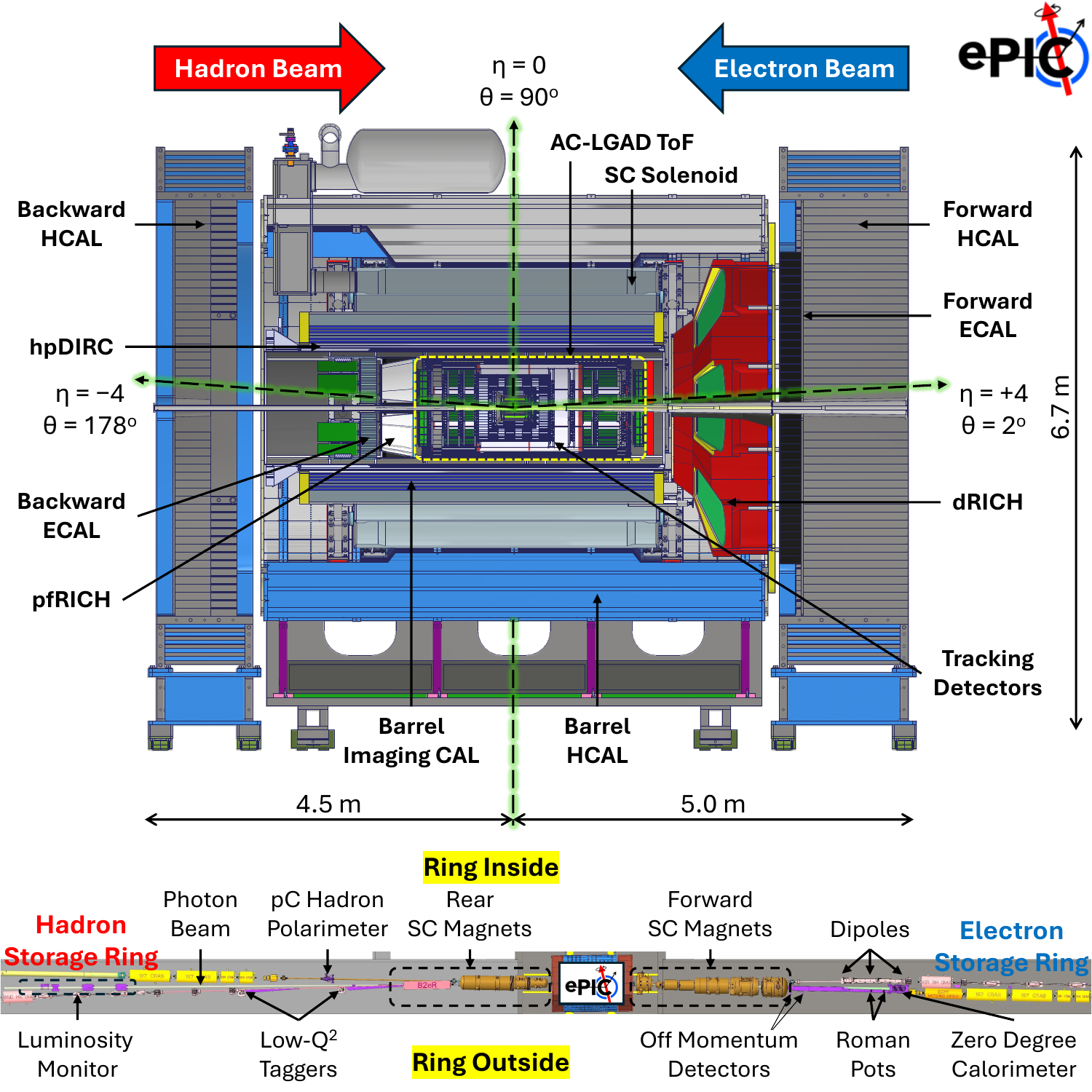}
\caption{\label{fig:ePIC_drawing}Schematic drawing of the ePIC detector (top) and 6 o'clock interaction region (bottom), where $\theta$ and $\eta$ are polar angle and pseudorapidity, respectively.}

\end{figure}

The Electron-Proton/Ion Collider (ePIC) detector~\cite{ePIC_web-page} will be installed in IR6, the 6-o’clock interaction region of the EIC (Fig.~\ref{fig:EIC_schematic_drawing}). It comprises vertex/tracking, particle-identification (PID), and calorimeter subsystems (Fig.~\ref{fig:ePIC_drawing}). 

The interaction point (IP) beam pipe has a \SI{62}{mm} inner diameter and extends \SI{80}{cm} (electron-going) and \SI{67}{cm} (hadron-going) from IP6. It is made of $\SI{757}{\upmu m}$ thick beryllium with a $\SI{5}{\upmu m}$ thick gold internal coating.

The silicon vertex and tracking system~\cite{LI2023168687}, based on the Monolithic Active Pixel Sensor (MAPS) technology, provides a low material budget ($\sim \SI{0.05}{\%}~X/X_{0}$ per layer) and high spatial resolution ($\SI{10}{\micro m}$ pitch). Central tracking uses micropattern gaseous detectors (MPGD) to improve angular resolution and enhance $\pi$/$K$ separation. PID relies on several complementary systems: the AC-LGAD Time-of-Flight (ToF) detector in the barrel and forward region~\cite{LI2023168687}; a high-performance DIRC (hpDIRC)~\cite{kalicy2022developinghighperformancedircdetector} in the central region; the  proximity-focusing RICH (pfRICH)~\cite{pfRICH_CDR} and dual-radiator RICH (dRICH)~\cite{VALLARINO2024168834} detectors in the forward and backward regions, respectively, provide extensive $\pi/K/p$ separation. Surrounding these systems, high-resolution electromagnetic (ECAL) and hadron (HCAL) calorimeters: ECALs~\cite{ECAL_EIC} identify scattered leptons and photons in deep-inelastic scattering (DIS) and semi-inclusive processes, while HCALs support jet reconstruction and DIS kinematics.

In addition to the central ePIC subsystems, distant auxiliary detectors (rear/electron-going: low-$Q^{2}$ silicon pixel taggers and luminosity monitor; forward/hadron-going: zero-degree calorimeter (ZDC), roman pots, and off-momentum detectors) located more than \SI{5}{m} from the interaction point enhance the overall measurement capabilities by detecting particles at extreme angles and providing independent luminosity monitoring.

\subsection{Beam Loss Mechanisms and Background Sources}
Beam losses in the ESR will result from a combination of fundamental beam-dynamics processes, interactions with the environment, and machine-related perturbations. Understanding these loss mechanisms is essential for designing an effective collimation system, protecting SC magnets and the ePIC detector, and ensuring stable, high-luminosity operation. The major sources of electron beam losses relevant to the EIC and discussed in the paper are summarized below.

\paragraph{Touschek Scattering.}
Touschek scattering is a single hard Coulomb scattering between two particles within the same beam bunch, transferring transverse momentum into the longitudinal plane and inducing large energy deviations. Particles that exceed the RF or momentum acceptance are lost, making Touschek scattering a dominant beam-lifetime limitation in lepton storage rings.

\paragraph{Beam-Gas Interactions.}
Elastic and inelastic scattering off residual gas molecules cause betatron oscillation growth or large energy deviations. These processes include Coulomb scattering and Bremsstrahlung, both contributing to losses in dispersive regions or at aperture restrictions.

\paragraph{Bethe-Heitler Scattering.}
Hard scattering processes of the form $e^{-}p \rightarrow e^{-}p\gamma$ result in electrons with large momentum offsets. These particles are lost rapidly downstream of the IP and form an important background component in the ePIC detector.

Beam-induced background particles transported into the ePIC detector generate secondary showers that can degrade the detector’s physics performance by increasing occupancies, confusing pattern recognition, and reducing track-reconstruction efficiency. The cumulative radiation dose and neutron fluence from these showers pose an additional long-term risk, as they can damage radiation-sensitive components, particularly silicon sensors and front-end electronics, potentially impacting detector longevity, calibration stability, and operational reliability.

Beyond detector impacts, uncontrolled beam losses also deposit energy in the SC magnets surrounding the IP. Shower particles striking the cold mass can lead to both localized energy deposition, potentially triggering quenches, and cumulative radiation dose to insulation materials such as epoxy resins. Studies at the LHC have systematically investigated beam-induced quench levels in SC magnets, demonstrating how localized losses can approach or exceed quench thresholds and thereby inform machine protection strategies~\cite{PhysRevSTAB.18.061002}. Operational experience at SuperKEKB further illustrates this challenge: beam-induced quenches have been observed in the SC magnets of the final-focus system during routine operation, underscoring the sensitivity of high-current lepton machines to halo and collision-induced losses~\cite{OHUCHI2022165930}.

In addition to transient effects, cumulative radiation dose may contribute to long-term degradation of insulation materials if losses are not properly controlled. While such limits are not expected to be reached under nominal operating conditions, they provide an important constraint on acceptable loss levels. In this context, the collimation system is designed to localize beam losses in the IR by placing collimators at locations optimized in phase with respect to the final-focus SC magnets, thereby reducing both peak energy deposition and integrated dose in these sensitive components.

Therefore, maintaining beam losses near the IR below the established limits on radiation dose, neutron fluence, and cold-mass heat deposition, is essential both for detector performance and for safe, reliable operation of the ESR SC magnet system.

For the collimation system design discussed in this paper, we focus on the Touschek and beam–gas loss components. According to SuperKEKB measurements~\cite{NATOCHII2023168550}, these mechanisms are expected to represent the primary sources of particle losses in the EIC ESR and therefore play a central role in determining the required collimation efficiency and protection strategy.

Several effects are not yet included, such as machine and alignment errors, beam–beam interactions, the ePIC solenoid field, crab cavities, and injection-related losses. Among these, machine imperfections (e.g., orbit distortions and optics perturbations) are expected to have the largest direct impact on collimation efficiency, as they can modify the effective machine aperture and increase leakage from the collimation region. Beam–beam effects may contribute to halo formation and diffusion during collisions, while the impact of the detector solenoid and crab cavities is expected to be more localized and largely correctable through optics tuning. Injection-related losses are expected to be localized in time and not representative of steady-state operation, as the ESR will operate in a swap-out injection scheme with on-axis, on-energy injection of fresh bunches. Off-normal injection scenarios, including injection failures and associated losses, are beyond the scope of this work and will be addressed in future studies in the context of defining the EIC machine protection system (MPS) requirements.

In addition to the mechanisms considered here, other processes such as radiative scattering, collective effects, and beam instabilities may contribute to beam losses in high-current operation. In particular, operational experience at SuperKEKB has demonstrated the occurrence of fast, transient loss events (``sudden beam losses'') associated with beam instabilities~\cite{Ikeda2023SuddenLoss}. Such losses are not governed by steady-state halo dynamics and may not be efficiently mitigated by collimation alone, instead requiring fast detection and response through the MPS~\cite{nomaru2026measurementsuddenbeamloss,YOSHIHARA2025170117}. These effects are not included in the present study and will be addressed in future work.

\section{\label{sec:CollimationSystemDesign}Collimation System Design}
The general design overview of the ESR lattice at all energies is discussed in Ref.~\cite{Marx_2023}. The current lattice v6.3.1~\cite{Nosochkov:2025vmb} used in the study includes compensation of non-linear chromaticity but does not include the ePIC detector solenoid field, crab cavities required for the crab crossing scheme, or machine errors and misalignments. 

Previous studies at electron storage rings have shown that beam losses are often effectively single-turn at the point of final loss~\cite{PhysRevAccelBeams.24.081001}. This emphasizes the importance of placing the collimation system upstream of the protected region and as close as possible to it, in order to intercept halo particles on trajectories leading to losses before they reach sensitive components.

Although beam losses from Touschek scattering and beam–gas interactions are generated throughout the ring, only a fraction of these particles reaches the interaction region. The ESR collimation system is therefore designed to intercept this population in phase space before it is lost in IR6.

The betatron collimation insertion has been integrated into the ESR lattice at IR4, selected as the closest upstream insertion region to IR6 with favorable optics conditions for halo interception. IR4 provides a suitable compromise: it is sufficiently close to IR6 to intercept halo particles contributing to downstream losses, while also being far enough upstream (approximately \SI{600}{m}) to reduce the probability of forward-scattered particles propagating directly into IR6. The collimator locations are further guided by the phase advance with respect to the limiting apertures in IR6, enabling efficient interception of the relevant phase-space trajectories. A more systematic optimization of phase relationships and potential leakage, as well as the impact of machine imperfections, will be addressed in future studies.

\begin{figure*}[htbp]
\centering
\includegraphics[width=\linewidth]{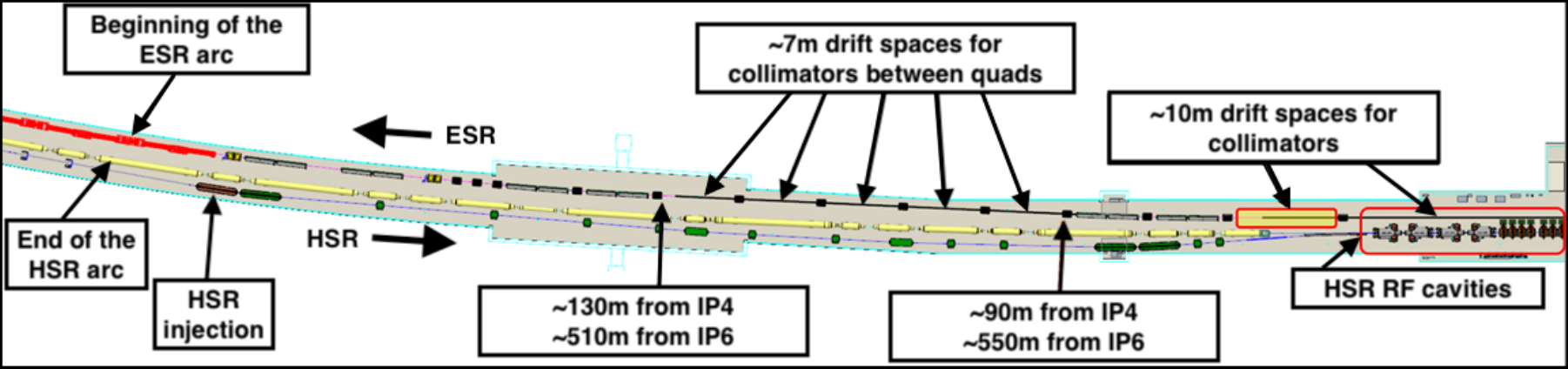}
\caption{\label{fig:ir4_lattice}Schematic layout of IR4 with emphasis on the drift regions reserved for collimator placement.}

\end{figure*}

\begin{figure*}[htbp]
\centering
\includegraphics[width=\linewidth]{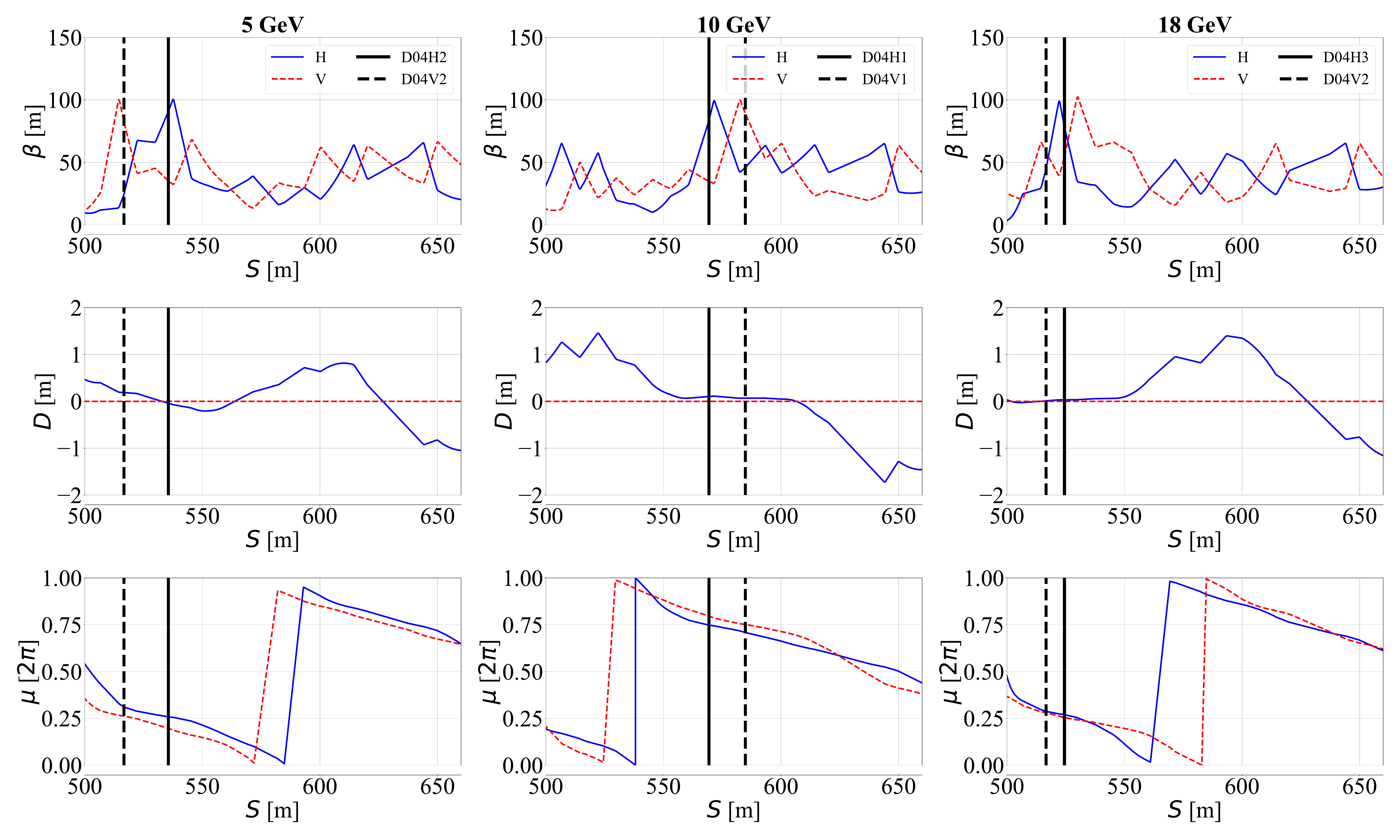}
\caption{\label{fig:ir4_optics}Optics functions of the ESR IR4 section in the collimation insertion. The beam direction is from right to left. Solid blue and red lines with markers represent the horizontal and vertical planes, respectively. Vertical dashed lines indicate the positions of the collimators relative to IP6, which is at $S = \SI{0}{m}$. The rows, from top to bottom, show the betatron functions, dispersion, and fractional part of the betatron phase advance (with respect to the IP6 phase), while the columns, from left to right, correspond to 5, 10, and \SI{18}{GeV} beam energies. Collimator names such as ``D04H1'' or ``D04V1'' indicate the first (``1'') horizontal (``H'') or vertical (``V'') collimator located in interaction region ``04''.}

\end{figure*}

Figure~\ref{fig:ir4_lattice} shows the schematic layout of the IR4 section designated for the collimation insertion. The presence of the injection insertions for both rings, as well as the HSR RF cavities located in IR4, makes it challenging to identify suitable positions for the ESR collimators without interfering with existing equipment or creating unnecessary beam losses near sensitive components. The available drift spaces for installing the collimators are located downstream of IP4, with most positioned after the HSR RF cavities. This placement minimizes the risk of scattered particles interacting with the injection hardware or RF structures.

For each collimator, a \SI{2}{m} drift space has been reserved. This choice is based on a preliminary mechanical design of the collimator chambers~\cite{ESR_Coll} inspired by those used at PEP-II~\cite{osti_877438} (SLAC, USA) and SuperKEKB~\cite{PhysRevAccelBeams.23.053501} -- electron-positron colliders with comparable circumference, beam energy, and beam current. These chambers, approximately \SI{1}{m} in active length, require additional drift on either side to accommodate tapers, diagnostics, and vacuum transitions.

The optics in the IR4 region have been carefully re-tuned to accommodate these collimator chambers while maximizing betatron collimation performance. In particular, the lattice has been arranged to provide:
\begin{itemize}
    \item large betatron functions (\SIrange{50}{100}{m}), enhancing the transverse amplitude of halo particles at the collimator jaw;
    \item zero dispersion, ensuring that collimation is dominated by betatron motion rather than momentum spread;
    \item a half-integer betatron phase advance with respect to the IR6 final-focusing quadrupoles, which have the smallest physical aperture in the ring due to the extremely large beta function of the order of $\SI{1}{km}$ required for strong beam focusing at the IP; and
    \item the overall phase advance over the entire insertion region has to meet certain requirements for dynamic aperture optimization.
\end{itemize}

This phase relationship ensures that particles intercepted at IR4 correspond to those most likely to impact the tight apertures in IR6 when generated upstream of IR4.

Figure~\ref{fig:ir4_optics} shows the resulting optics functions for the three studied beam energies with collimator locations, and Table~\ref{tab:Collimator_Optics_Parameters} summarizes the key collimator optics parameters. At each beam energy, the system uses only one horizontal and one vertical primary collimator, with possibility for the remaining collimators to be held at wider apertures to provide auxiliary halo cleaning when needed. The locations of the active primary collimators are chosen to coincide with optics conditions that maximize interception efficiency, as illustrated in the figure.

In addition to defining the nominal collimation settings, the presence of the additional collimators provides operational flexibility and margin. In particular, they can intercept halo particles at larger amplitudes and help mitigate potential increases in beam losses arising from undesirable effects such as optics perturbations, orbit distortions, or alignment errors. This approach enables adjustment of the collimation hierarchy during commissioning and operation without introducing additional primary collimators, while preserving the efficiency of the baseline configuration. A similar operational strategy has been successfully employed at SuperKEKB~\cite{Ishibashi:2017,PhysRevAccelBeams.23.053501}, where a limited set of primary collimators defines the core cleaning performance, while additional collimators are progressively activated and optimized during commissioning and luminosity ramp-up to accommodate increasing beam current and evolving machine conditions.

\begin{table}[htbp]
    \caption{\label{tab:Collimator_Optics_Parameters}
    Optics parameters and settings of the primary collimators for different beam energies. Here, $S$ denotes the distance from IP6, $\beta$ the betatron function, $\sigma_{\beta} = \sqrt{\beta \varepsilon}$ the RMS transverse beam size due to betatron motion, $\varepsilon$ the RMS emittance, $D$ the dispersion, and $\Delta\mu$ the betatron phase advance relative to the location of the narrowest aperture in the IR6 final focusing quadrupoles ($A_\mathrm{IR6}$). The collimator aperture (half-opening) is indicated as $A_\mathrm{coll.}$. Values are given separately for the horizontal and vertical planes, corresponding to the horizontal and vertical collimators, respectively.}
    \centering
    \begin{tabular}{c|c|c|c|c|c|c|c}
        \hline\hline
        Energy & Collimator & $S$ & $\beta$ & $D$ & $\Delta\mu$ & $A_\mathrm{coll.}$ & $A_\mathrm{IR6}$ \\ 
        \multicolumn{1}{c|}{[GeV]} & \multicolumn{1}{c|}{Name} & \multicolumn{1}{c|}{[m]} & \multicolumn{1}{c|}{[m]} & \multicolumn{1}{c|}{[cm]} & \multicolumn{1}{c|}{[$2\pi$]} & \multicolumn{1}{c|}{[$\sigma_{\beta}$] ([mm])} & \multicolumn{1}{c}{[$\sigma_{\beta}$]}\\
        \hline
        \multirow{2}{*}{5}  & D04H2 & 535.53 & 90.84 & -4.00 & 0.51 & 15 (20.2) & 19\\
                             & D04V2 & 516.63 & 82.20 & 0.01  & 0.44 & 21 (8.1) & 27\\
        \hline
        \multirow{2}{*}{10} & D04H1 & 569.33 & 83.41 & 10.31 & 0.01 & 9 (11.6) & 15\\
                             & D04V1 & 584.83 & 88.99 & -0.02 & 0.04 & 21 (7.1) & 28\\
        \hline
        \multirow{2}{*}{18} & D04H3 & 524.43 & 78.19 & 3.31  & 0.48 & 12 (16.4) & 15\\
                             & D04V2 & 516.63 & 58.03 & 0.00  & 0.46 & 18 (6.1) & 23\\
        \hline\hline
    \end{tabular}
\end{table}

\section{\label{sec:MultiTurnParticleTracking}Multi-Turn Particle Tracking}
Below we describe the methodology for beam loss process simulation and particle tracking through the machine element sequence.

\subsection{Simulation Framework}
We developed an EIC-specific multi-turn particle-tracking tool based on the Xsuite framework~\cite{Iadarola:2023fuk}, which has also been used in recent studies of the FCC-ee collimation system design, including detailed investigations of loss dynamics and collimation performance~\cite{Broggi:2026collimation} -- a modern, modular simulation environment that enables symplectic particle dynamics modeling together with a broad range of beam environment interaction effects, including synchrotron radiation, space charge, electron cloud, and beam-beam interactions. For collimation studies, Xsuite provides an interface to BDSIM~\cite{NEVAY2020107200}/Geant4~\cite{AGOSTINELLI2003250,1610988,ALLISON2016186}, allowing detailed simulation of particle-matter interactions inside collimator materials.

Although Xsuite includes many beam-physics processes and provides interfaces for modeling beam–matter interactions (e.g., collimators, beam–gas scattering, and beam–beam interactions), at the time this study was performed it did not include fully integrated native implementations of all scattering mechanisms required for realistic ESR beam loss simulations, in particular for Touschek scattering. These processes are machine-specific and typically require dedicated modeling. We note that native implementations of such processes are under active development and are expected to be included in future releases of Xsuite. To address this, we implemented beam–gas and Touschek scattering models tailored to the ESR, following the approach developed for SuperKEKB~\cite{PhysRevAccelBeams.24.081001, NATOCHII2023168550}, as summarized below.

After loading the ESR lattice and completing optics initialization -- including aperture definitions and RF voltages of 12, 23.7, and \SI{68}{MV} for 5, 10, and \SI{18}{GeV} beams, respectively, distributed across 18 RF cavities -- the collimators are instantiated as ``Geant4Collimator'' elements. Each collimator jaw is modeled as a \SI{5}{cm}-long (along the beam axis) tungsten block, corresponding to approximately 14 radiation lengths (RL). This choice is motivated by operational experience at other lepton machines and studies of collimator interception efficiency. At the Large Electron-Positron (LEP) collider~\cite{vonHoltey1987LEPCollimation}, tungsten collimators with lengths of order \SI{30}{RL} were used at beam energies around \SI{50}{GeV}, while at SuperKEKB (\SIrange{4}{7}{GeV})~\cite{PhysRevAccelBeams.23.053501}, a thickness of order \SI{3}{RL} has been shown to be sufficient to induce large angular deflections and remove particles from the machine acceptance. Consistent with these observations, our preliminary simulations (to be reported separately) indicate that a thickness of approximately \SI{1}{RL} is sufficient to achieve the required reduction of beam losses in IR6, with no significant improvement for larger thicknesses. The use of \SI{5}{cm} of tungsten therefore provides a conservative approximation of a near-ideal absorber, ensuring efficient particle removal and enabling an upper bound estimate of collimation performance. Further optimization of the material and geometry will be addressed in future engineering studies.

The typical high-statistics Touschek and beam-gas tracking simulations begin by defining approximately 1000 discrete scattering points located at every quadrupole magnet and extra points at \SI{10}{cm} intervals from \SI{-50}{m} (electron-going) to $+\SI{35}{m}$ (hadron-going) around IP6. At each scattering location, macro-particles are generated inside the 3D Gaussian bunch volume, where each macro-particle represents a group of physical particles through an assigned weight. Their initial momenta, energies, and statistical/scattering weights are determined by the underlying physics processes~\cite{NATOCHII2023168550}:
\begin{itemize}
    \item Coulomb (elastic) scattering: modeled using Rutherford's formula with a cutoff Coulomb potential and a screening effect to regularize small-angle scattering~\cite{Jackson1962,ChaoTigner1999}.

    \item Bremsstrahlung (inelastic): implemented according to Bethe-Heitler theory using the Koch-Motz formalism for complete screening in the Born approximation~\cite{BetheHeitler1934,KochMotz1959}.

    \item Touschek scattering: based on M\o{}ller's non-relativistic differential cross-section~\cite{Moeller1932}, combined with Bruck's formula~\cite{Bruck1974} for calculating the local loss rate. 
\end{itemize}
The scattering weights normalization procedure includes computing absolute loss rates by scaling each macro-particle to the local beam-gas and Touschek scattering probabilities. All loss rates quoted in this work correspond to the total stored beam.

All primary electrons and any secondary particles produced in the collimators with kinetic energy above \SI{10}{\%} of the nominal ESR beam energy are tracked for the purposes of the loss-pattern and lifetime study. This threshold was chosen as a conservative upper bound, since particles with relative energy deviations exceeding approximately $10 \times \Delta p/p \approx \SI{1}{\%}$ are already outside the momentum acceptance of the ring, as discussed later in Section~\ref{sec:MachineAcceptance}, and tracking particles with substantially larger energy deviations would not be meaningful for the objectives of this study while significantly increasing computational cost. Lower-energy shower particles, which are not relevant for the beam loss statistics presented here, will be included in future dedicated radiation and heat load simulations. The tracking continues until particles are intercepted by the machine aperture, absorbed in a collimator, or complete 200~turns. Synchrotron radiation damping and excitation, as well as RF (de)acceleration, are included throughout the tracking. We verified that extending the tracking to a larger number of turns does not alter the loss distribution or lifetime estimates within statistical uncertainty, and the loss rates after 200~turns drop by 3-4 orders of magnitude as shown in Section~\ref{sec:BeamLossLifetimeEstimates}. This conclusion is specific to the present simulation setup, which assumes ideal optics conditions and includes only Touschek and beam–gas scattering processes, without magnetic errors, beam–beam effects, or other additional loss mechanisms.

The coordinates and properties of all lost particles are recorded, forming the basis for beam losses, lifetime, and collimation efficiency analyses.

\subsection{ESR Vacuum}

The beam-gas scattering weights described above are based on a realistic residual gas pressure profile estimated using the Synrad+/Molflow+ framework~\cite{Kersevan:2019vme}, which traces synchrotron radiation photons and calculates their flux and power distribution on the inner surface of the beam pipe.

\begin{figure}[htbp]
\centering
\includegraphics[width=\linewidth]{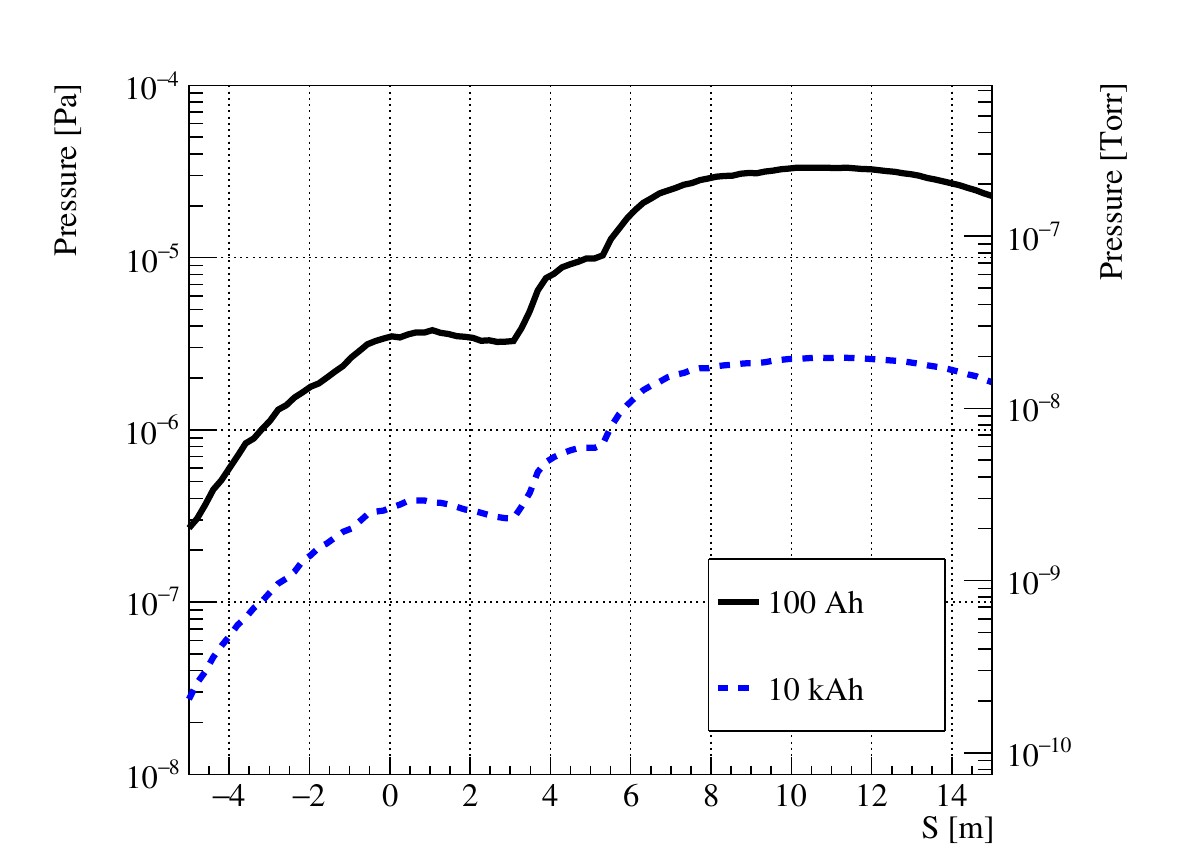}
\caption{\label{fig:ESR_Pressure}
Estimated residual $H_{2}$ gas pressure in ESR IR6 for delivered beam doses of \SI{100}{Ah} (black solid line) and \SI{10}{kAh} (blue dashed line). The beam direction is from right to
left. The corresponding average pressures are \SI{15.4}{\upmu Pa} (\SI{115.2}{nTorr}) and \SI{1.3}{\upmu Pa} (\SI{9.4}{nTorr}), respectively.}

\end{figure}

For the studies presented in this paper, we assumed a vacuum corresponding to \SI{10}{kAh} of delivered beam dose, resulting in the gas pressure profile shown in Fig.~\ref{fig:ESR_Pressure}, predominantly composed of molecular hydrogen ($H_{2}$). The contributions of other residual gas species such as $N_{2}$, $CO$, $CO_{2}$, and $CH_{4}$, which are typically present only at very low concentrations in similar ultra-high-vacuum environments, are considered negligible for this study.

The vacuum model assumes photon-stimulated desorption behavior consistent with synchrotron-radiation-dominated storage rings, where the desorption yield decreases with accumulated photon dose due to surface conditioning. This approach follows the EIC CDR vacuum design~\cite{osti_1765663}, which is based on empirical performance observed in existing facilities, and therefore reflects expected steady-state operating conditions rather than initial commissioning.

Since the installation of vacuum pumps in the inner section of IR6 (\SI{-5}{m} to $+\SI{15}{m}$) is challenging, the vacuum profile was modeled in detail for this specific region. Outside of this section, a uniform pressure of \SI{5}{nTorr} was assumed around the rest of the ring.

While the vacuum profile is modeled in detail in IR6 due to its direct impact on detector backgrounds, a more comprehensive treatment of the vacuum conditions in the collimation region (IR4) and their influence on local loss generation will be addressed in future studies as the ESR vacuum system design matures.

\section{\label{sec:TrackingSimulationResults}Tracking Simulation Results}
The collimation simulations are based on the ESR lattice optics, physical beam pipe aperture, and the limiting apertures in the ring, including detailed IR6 apertures, which define the primary constraints for beam loss localization. Collimator settings are determined through aperture scans, in which the collimator half-opening is varied relative to the local beam size and compared to the effective IR6 aperture scaled by the corresponding optics functions. This provides an analytical reference linking collimator settings to the machine acceptance. In this framework, the optimal collimator settings are identified as those that significantly reduce IR6 losses while preserving beam lifetime.

Beam losses are evaluated through multi-turn tracking of particles generated from Touschek and beam–gas scattering processes at distributed locations around the ring. Particles are followed until they are either intercepted by collimators or lost at the machine aperture, allowing a self-consistent determination of loss locations and collimation efficiency. Interactions of particles with the collimator material, including the production of secondary particles, are simulated using the Xsuite–BDSIM interface. The simulations are performed under ideal optics conditions, without inclusion of machine errors or orbit distortions.

In this section, we present the results of the multi-turn particle tracking in the ESR lattice, including collimator aperture scans, machine acceptance estimates, and beam loss distributions around the ring at different beam energies.

\subsection{Collimator Scan}
To determine the optimal aperture for each collimator, we performed a scan of the collimator half-opening (i.e., the distance between the jaw and the beam center), evaluating both the beam lifetime and the local losses in the IR6 cryostat region (\SI{-15}{m} to $+\SI{22.5}{m}$). The simulations are performed with fixed collimator settings for each scan point, and particles are generated from scattering processes distributed around the ring. The results therefore represent steady-state loss conditions, with no transient loss spikes associated with collimator movement.

\begin{figure}[htbp]
\centering
\subfloat[\label{fig:ESR_lifetime_collScan_10GeV}Beam lifetime.]{\includegraphics[width=\linewidth]{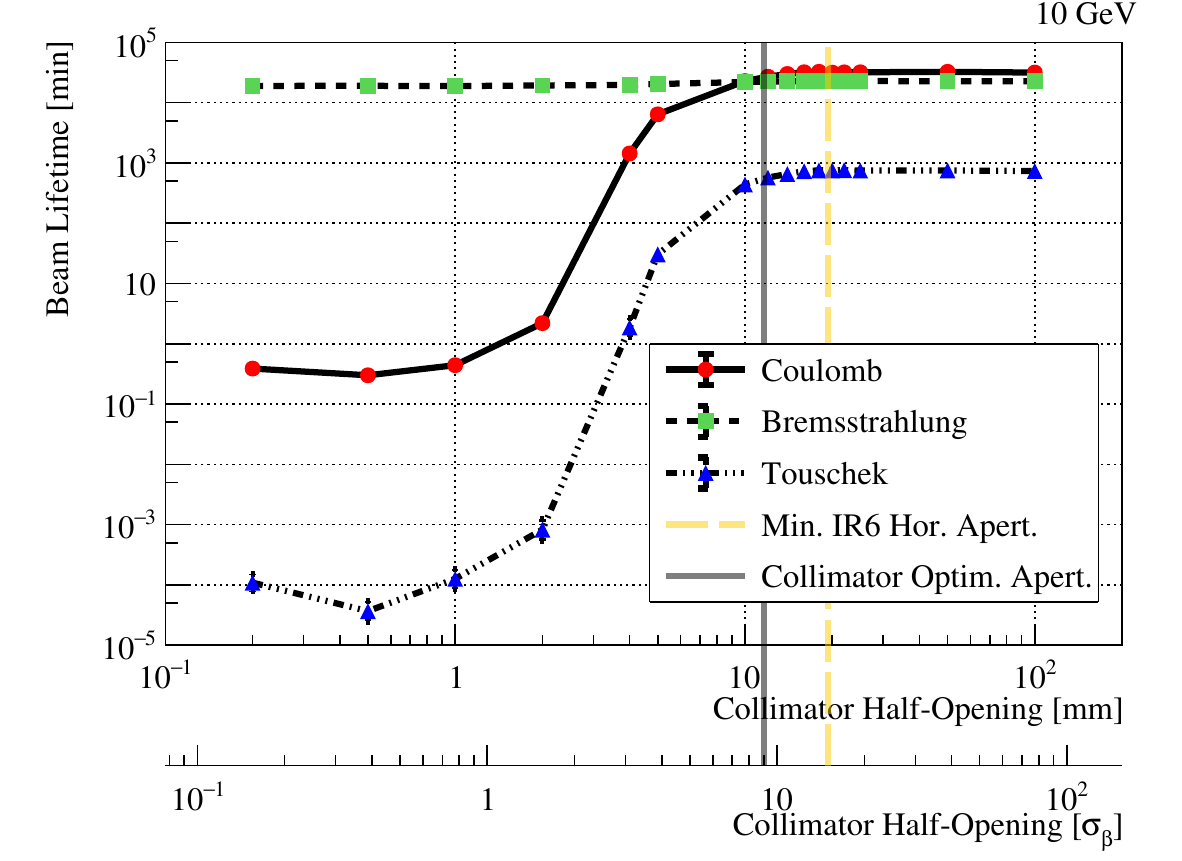}}
\\
\subfloat[\label{fig:ESR_IR6losses_collScan_10GeV}Beam losses.]{\includegraphics[width=\linewidth]{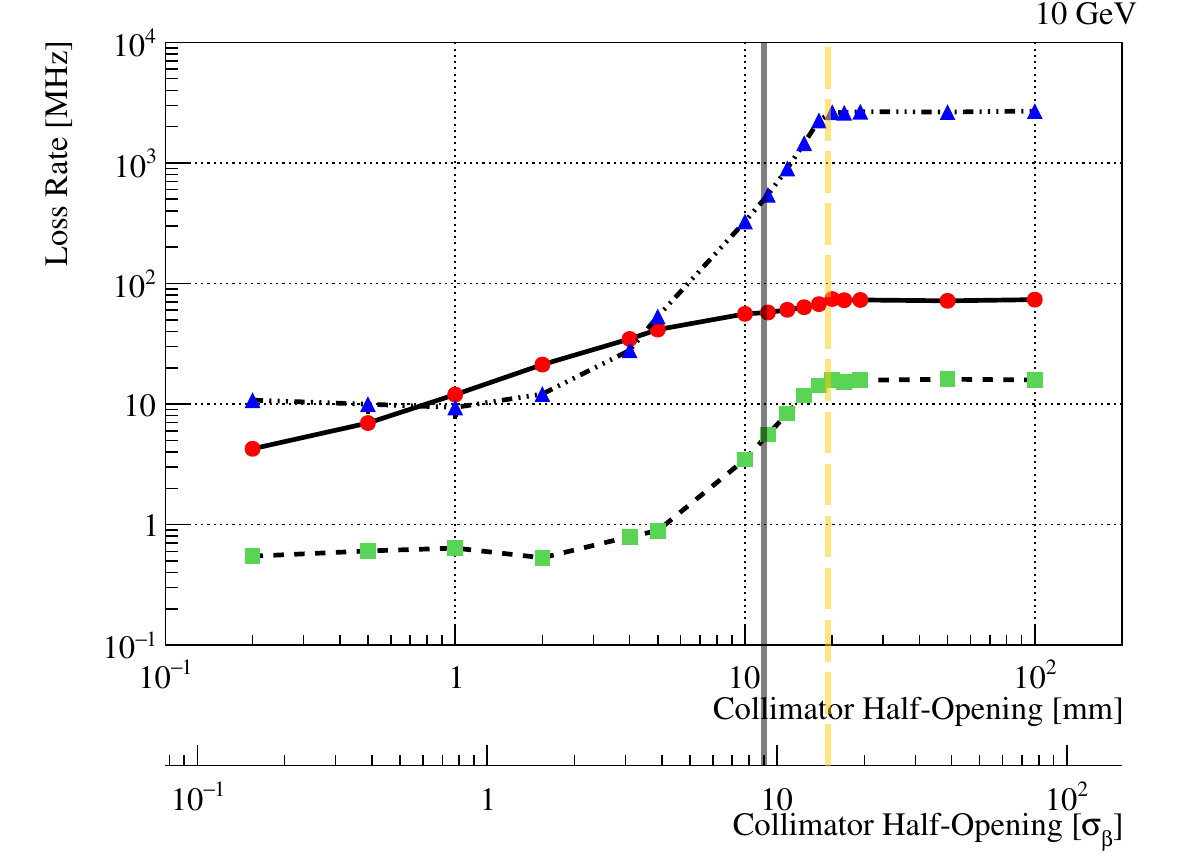}}
\caption{\label{fig:CollApertScan}
ESR beam lifetime and IR6 losses as a function of the D04H1 collimator aperture at \SI{10}{GeV}. The IR magnet aperture, indicated by the orange dashed vertical line, is scaled relative to the collimator aperture by the square-root ratio of the corresponding betatron functions.}

\end{figure}

Figure~\ref{fig:CollApertScan} presents an example of the simulated loss rates for Coulomb (red circles), Bremsstrahlung (green squares), and Touschek (blue triangles) processes at \SI{10}{GeV}. The minimum IR6 aperture within the final focusing system is indicated by the vertical orange dashed line. As expected, adjusting the collimator aperture above this IR6 limit has no noticeable impact on the overall beam losses. In contrast, narrowing the collimator results in a substantial reduction of losses in the IR6 cryostat, accompanied by a decrease in beam lifetime due to increased halo interception by the collimator rather than the IR6 beam pipe. The optimal collimator aperture (vertical grey solid line) is therefore defined as the setting at which the beam lifetime remains essentially unchanged while the IR6 losses are significantly suppressed. The corresponding aperture values for all studied beam energies are summarized in Table~\ref{tab:Collimator_Optics_Parameters}. These optimal settings are found to be consistent with the expected scaling based on the IR6 limiting aperture and local optics functions, confirming that the simulation results are in good agreement with the analytical aperture estimates.

Beam–beam effects may contribute to halo formation and diffusion during collisions, potentially increasing the particle flux intercepted by the collimation system, while machine imperfections can modify the effective aperture and phase relationships and thereby affect collimation efficiency; their quantitative impact will be addressed in future studies.

\subsection{\label{sec:MachineAcceptance}Machine Acceptance}
To evaluate the impact of the collimation system on the machine acceptance, defined by both the dynamic aperture\footnote{``The dynamic aperture (DA) is defined as the maximum phase-space amplitude within which particles do not get lost as a consequence of single-particle-dynamics effects.''~\cite{ChaoTigner1999}} and the physical beam pipe, we performed particle tracking studies scanning the transverse and momentum phase-space coordinates. Particles were launched at IP6 and tracked for 200~turns with RF cavities and synchrotron radiation enabled. The machine acceptance is defined as the region in phase-space for which particles remain stable (i.e., not lost). Random and systematic machine errors (e.g., magnet misalignments, magnetic field imperfections) are not included. Increasing the number of simulated turns or adjusting the stability criterion does not significantly affect the final results.

\begin{figure*}[htbp]
\centering
\subfloat[\label{fig:MA_xdp_plot}Momentum versus horizontal machine acceptance for the electron ring with and without collimators.]{\includegraphics[width=0.5\linewidth]{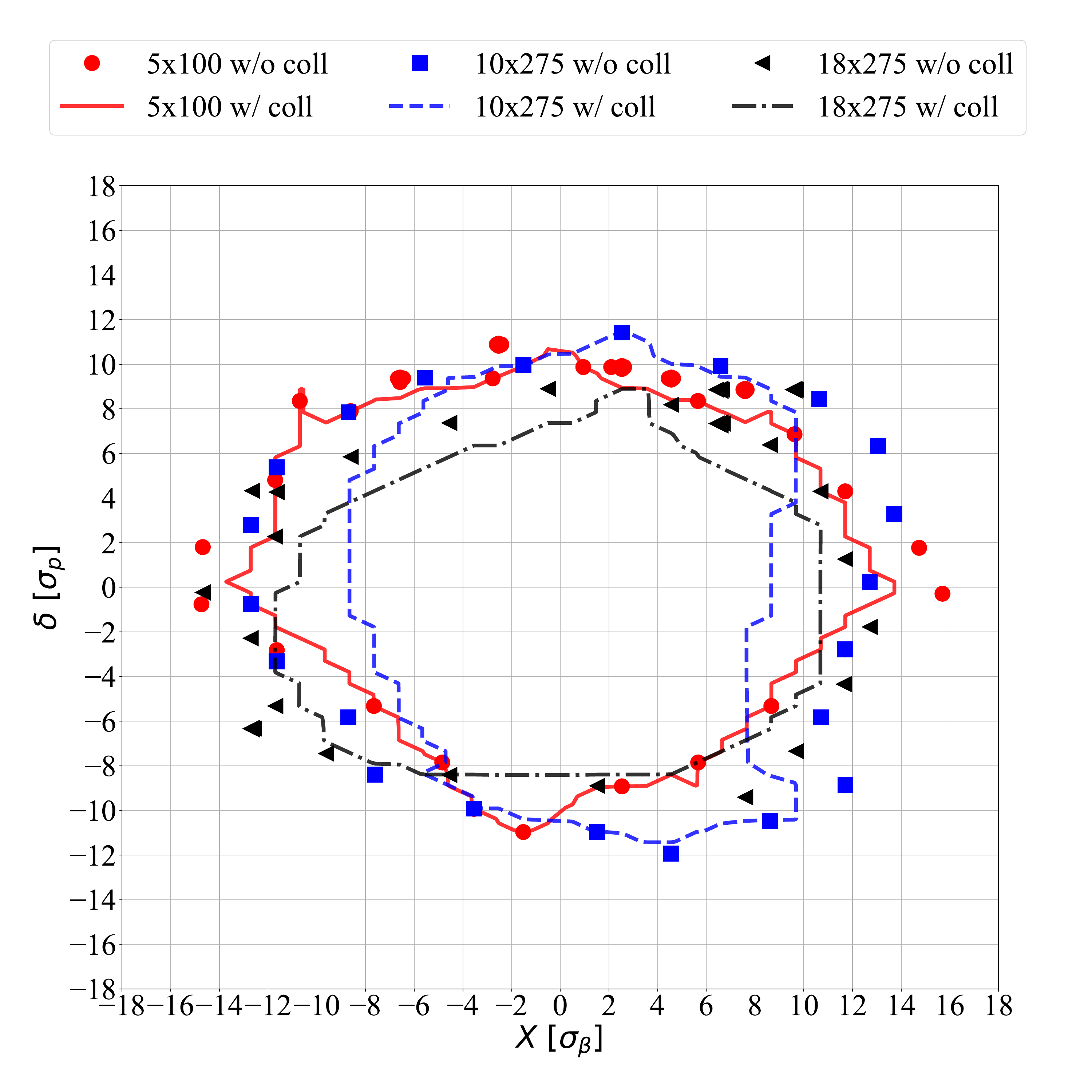}}
\hfill
\subfloat[\label{fig:MA_xy_plot}Transverse machine acceptance at $\delta = 0$.]{\includegraphics[width=0.5\linewidth]{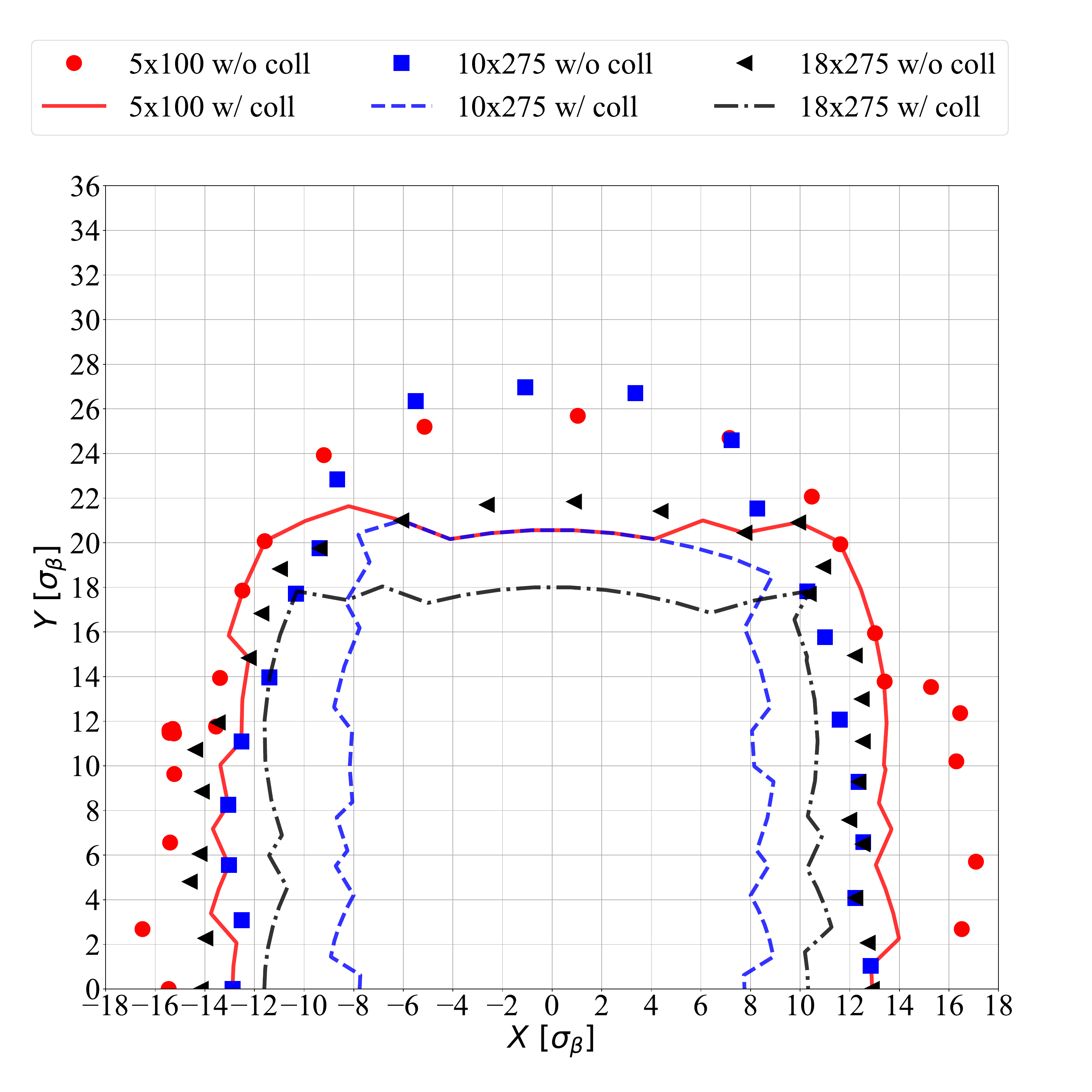}}
\caption{\label{fig:MachineAcceptance}
Machine acceptance in the presence and absence of collimators, where $\delta = \Delta p/p$ denotes the momentum offset. Markers and line styles distinguish the configurations with (``w/ coll'') and without (``w/o coll'') collimators.
}

\end{figure*}

Figure~\ref{fig:MachineAcceptance} presents the results of the machine acceptance simulations. Red circles (solid), blue squares (dashed), and black rectangles (dash-dotted) correspond to the cases without (with) collimators at 5, 10, and \SI{18}{GeV}, respectively.

The momentum acceptance at all three energies is $\SIrange{8}{10}{\sigma_{p}}$, consistent with the requirements specified in the EIC Conceptual Design Report (CDR)~\cite{osti_1765663}. The introduction of collimators primarily reduces the transverse acceptance due to betatron scraping in a zero-dispersion insertion. This reduction is on the order of $\SIrange{1}{3}{\sigma_{\beta}}$, which does not significantly impact the beam lifetime, keeping it at the multi-hour level as discussed later in the text.

\subsection{\label{sec:BeamLossLifetimeEstimates}Beam Loss and Lifetime Estimates}
To evaluate the performance of the collimation system and quantify beam losses around the ring, we performed high-statistics tracking simulations of scattered macro-particles. For each beam loss scenario, statistically converged multi-turn tracking simulations were performed by sampling approximately $6.5 \times 10^{7}$ macro-particles at all scattering locations and propagating them through the lattice.

\begin{figure*}[htbp]
\centering
\subfloat[\label{fig:ring_beamlosses_rate}Total (beam-gas + Touschek) beam losses around the ring without (black) and with (red) collimators. The location of six IPs is indicated with labels. Blue markers show beam losses at the collimators.]{\includegraphics[width=\linewidth]{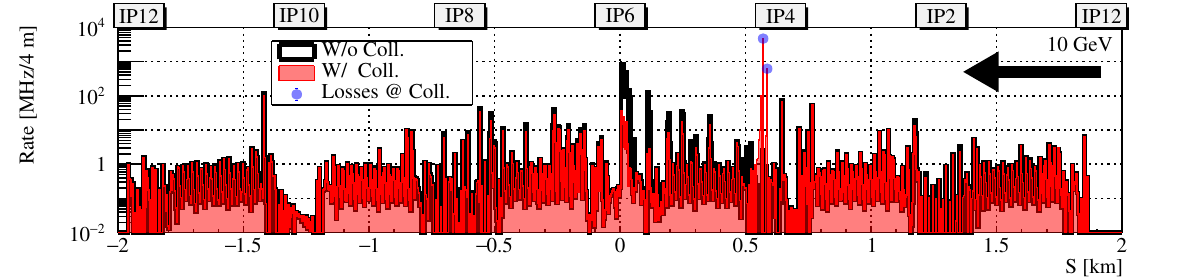}}
\\
\subfloat[\label{fig:ring_beamlosses_rate_ratio}Ratio of total (beam-gas + Touschek) beam losses around the ring between configurations without (black) and with (red) collimators.]{\includegraphics[width=\linewidth]{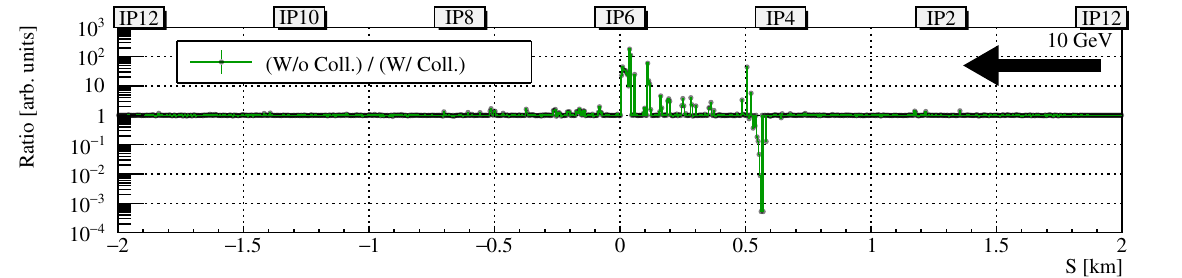}}
\\
\subfloat[\label{fig:ir6_beamlosses_rate}Beam losses in the IR6 final focusing cryostat region with closed collimators installed in the ring.]{\includegraphics[width=\linewidth]{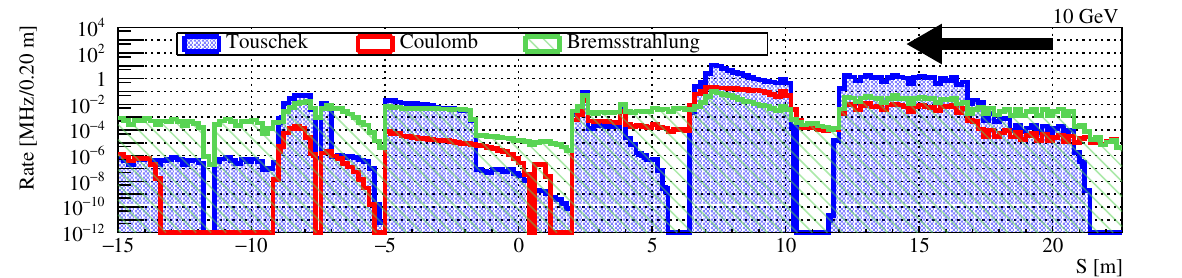}}
\\
\subfloat[\label{fig:ir6_beam_pipe_optics}The IR6 beam pipe aperture and beam envelope. The IP beam pipe and SC final focusing quadrupole sections are highlighted with vertical bands.]{\includegraphics[width=\linewidth]{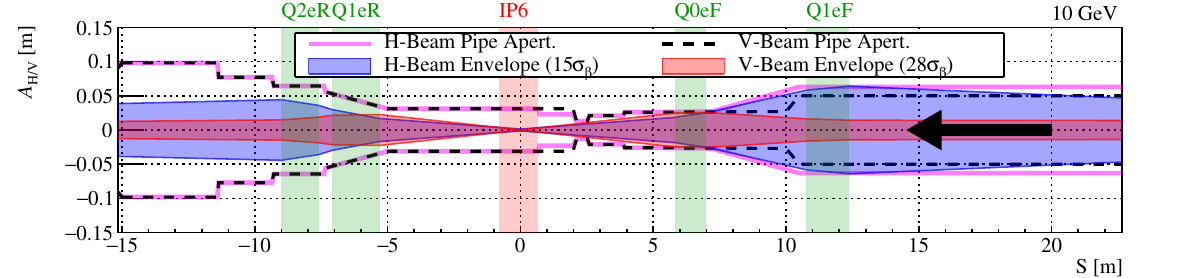}}
\caption{\label{fig:ESR_Beam_Losses}
Simulation results of ESR beam losses in IR6 at \SI{10}{GeV}. The ESR beam direction is indicated by the arrow.
}

\end{figure*}

Figure~\ref{fig:ESR_Beam_Losses} summarizes the simulated beam loss patterns around the ring (top), in the IR6 cryostat and ePIC detector region (middle), and the IR6 beam pipe aperture with the corresponding beam envelope (bottom). Outside the IR6 region, the beam pipe is modeled using a uniform geometry representative of the ESR arcs, namely an octagonal cross-section with a horizontal aperture of \SI{80}{mm} and a vertical aperture of \SI{36}{mm}, consistent with the baseline vacuum chamber design~\cite{Hetzel2025EICVacuum}.

From Fig.~\ref{fig:ring_beamlosses_rate}, one observes that without collimators the losses are distributed nearly uniformly across the ESR arcs, with a pronounced peak in IR6 due to its tight aperture in units of $\sigma_{\beta}$. Installing two primary collimators in IR4 reduces the IR6 losses by roughly two orders of magnitude (Fig.~\ref{fig:ring_beamlosses_rate_ratio}). As shown in Fig.~\ref{fig:ir6_beamlosses_rate}, losses in IR6 can reach up to \SI{10}{MHz} per \SI{20}{cm} with closed collimators installed in the ring, creating ``hot spots'' that generate heat-load densities\footnote{The power density is obtained from the kinetic energy loss rate $R_\mathrm{E}$ using $P[\mathrm{mW}] = R_\mathrm{E}[\mathrm{eV/s}]\cdot (1.602\times 10^{-16})$, assuming full local deposition, and then normalized by the longitudinal bin size (mW/cm).} of order \SI{1}{mW/cm} hitting the inner surface of the beam pipe in the cryostat region. With the collimation system in place, the attenuated heat load after passing the beam pipe materials remains below the \SI{3}{mW/cm} limit (derived from an assumed allowable heat load of about \SI{5}{W} per \SI{16}{m} cryostat cold mass section, consistent with preliminary EIC cryogenic design studies) -- an important result that will be discussed in detail in forthcoming publications on radiation and heat loads in the SC magnets.

A localized loss peak observed at approximately $\SI{-1.4}{km}$ (Fig.~\ref{fig:ring_beamlosses_rate}) originates primarily from Touschek-scattered particles that pass through the RF section in IR10, where they gain energy and are driven outside the machine momentum acceptance, and are subsequently lost in high-dispersion regions at the entrance of the downstream arc. This feature is present both with and without collimation and is only moderately affected by the collimation system, as it is dominated by off-momentum losses.

\begin{figure}[htbp]
\centering
\includegraphics[width=\linewidth]{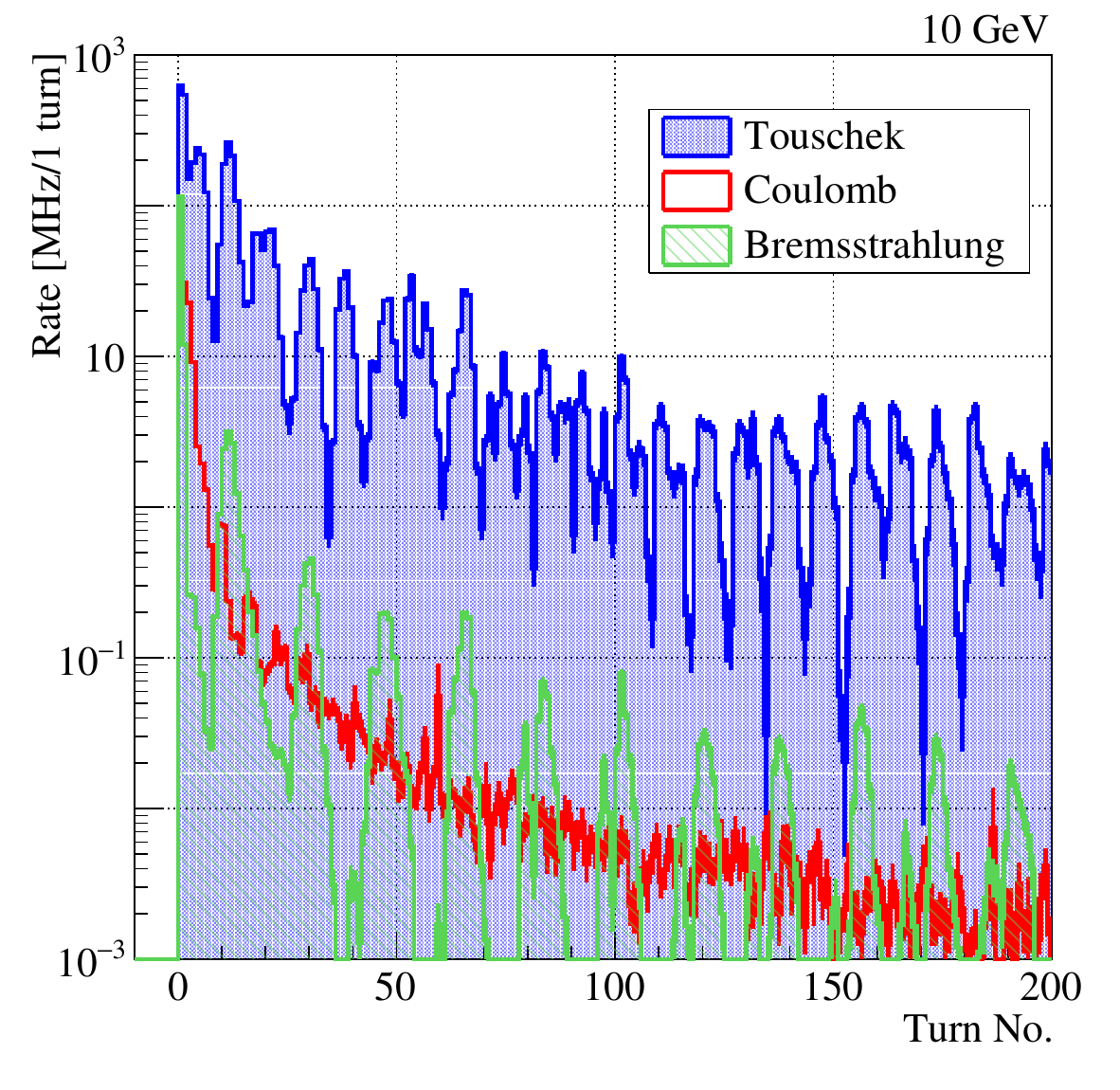}
\caption{\label{fig:ring_loss_timing_wo_coll}
Distribution of the turn at which the macro-particles are lost after been scattered due to Touschek (blue), Coulomb (red), or Bremsstrahlung (green) processes at \SI{10}{GeV}.}

\end{figure}

Figure~\ref{fig:ring_loss_timing_wo_coll} presents an example of the time distribution of beam losses at \SI{10}{GeV} without collimators. The majority of losses occur within a single turn, highlighting the need to place collimators as close as possible to IR6 because most stray particles reach the IR6 aperture in one revolution. The loss rates decrease by up to four orders of magnitude within the first 200~turns, indicating that additional tracking turns would not contribute meaningfully to the beam loss evaluation discussed in this paper.

\begin{table*}[htbp]
\centering
    \caption{\label{tab:LossesLifetime}
    Estimated IR6 beam losses, beam lifetime, and collimation system local cleaning inefficiency.}
    \begin{tabular}{l|c|c|c|c|c|c}
    \hline\hline
    \multirow{3}{*}{Energy [GeV]} & \multirow{3}{*}{Processes} & \multicolumn{2}{c|}{Without Collimators} & \multicolumn{3}{c}{With Collimators} \\
    \cline{3-7}
                        & & IR6 Losses & Lifetime & IR6 Losses & Lifetime & IR6 Cleaning \\
                        & & [MHz] & [hours] & [MHz] & [hours] & Inefficiency $\mathrm{[10^{-6}/m]}$\\
    \hline
    \multirow{3}{*}{5}  & Touschek & $7428.2 \pm 22.2$ & $2.7 \pm 0.1$ & $52.4 \pm 1.1$ & $2.6 \pm <0.1$ & \multirow{3}{*}{$88.5 \pm 1.6$}\\
                        & Coulomb & $179.7 \pm 0.6$ & $194.9 \pm 0.4$ & $7.9 \pm 0.1$ & $128.1 \pm 0.3$\\
                        & Bremsstrahlung & $10.5 \pm 0.1$ & $385.5 \pm 0.3$ & $0.7 \pm <0.1$ & $384.9 \pm 0.3$\\
    \hline
    \multirow{3}{*}{10} & Touschek & $2357.9 \pm 7.8$ & $12.7 \pm <0.1$ & $74.6 \pm 0.9$ & $9.1 \pm <0.1$ & \multirow{3}{*}{$395.9 \pm 4.7$}\\
                        & Coulomb & $62.7 \pm 0.2$ & $526.4 \pm 1.3$ & $2.5 \pm 0.1$ & $301.0 \pm 0.9$\\
                        & Bremsstrahlung & $13.5 \pm 0.1$ & $382.7 \pm 0.3$ & $1.5 \pm 0.1$ & $372.9 \pm 0.3$\\
    \hline
    \multirow{3}{*}{18} & Touschek & $2.2 \pm <0.1$ & $799.0 \pm 2.2$ & $0.2 \pm <0.1$ & $572.0 \pm 1.9$ & \multirow{3}{*}{$686.6 \pm 5.1$}\\
                        & Coulomb & $1.5 \pm <0.1$ & $2006.1 \pm 5.4$ & $0.1 \pm <0.1$ & $1316.1 \pm 4.1$\\
                        & Bremsstrahlung & $0.7 \pm <0.1$ & $432.1 \pm 0.3$ & $0.1 \pm <0.1$ & $416.2 \pm 0.3$\\
    \hline\hline
    \end{tabular}
\end{table*}
  
Table~\ref{tab:LossesLifetime} summarizes the beam loss rates in IR6 and the corresponding beam lifetime estimates due to beam-gas and Touschek processes at 5, 10, and \SI{18}{GeV}. At optimal aperture settings, the IR4 collimators reduce the IR6 loss rate by one to two orders of magnitude at all energies while leaving the beam lifetime essentially unchanged.

Because the ESR will operate with a swap-out injection scheme, all bunches in the ring will be replaced over approximately $T_{\mathrm{inj}} \approx \SI{20}{min}$ for 1160 bunches at a \SI{1}{Hz} injection rate. The shortest estimated Touschek lifetime, $\tau_{\mathrm{Touschek}} = \SI{2.6}{h}$ at \SI{5}{GeV}, results in an individual bunch-intensity reduction of $1 - \exp(-T_{\mathrm{inj}}/\tau_{\mathrm{Touschek}}) \approx \SI{12}{\%}$, keeping the average beam current at roughly the \SI{94}{\%} level -- acceptable for EIC operations.

Nevertheless, the beam lifetime remains an important parameter, as the gradual intensity decay between injections leads to transient bunch-to-bunch variations that can affect beam–beam interactions, luminosity stability, and background conditions at the interaction point.

Higher beam losses and correspondingly shorter lifetimes are expected at lower energies due to the Touschek loss scaling with the Lorentz factor $\gamma$, bunch intensity $N$, and momentum acceptance $\hat{\Delta p}/p \sim \SIrange{8}{10}{\sigma_p}$ (Fig.~\ref{fig:MA_xdp_plot})~\cite{Lee2019}:
\begin{equation}
    \tau \sim \frac{\gamma^2}{N}\left( \frac{\hat{\Delta p}}{p} \right)^3.
\end{equation}

At \SI{5}{GeV}, the beam has the lowest energy, leading to the shortest Touschek lifetime and highest loss rate. Because the gas pressure in IR6 is higher than the ring average pressure assumed in the simulations, IR6 contributes the dominant fraction of beam-gas losses. This suggests that improved vacuum conditions in IR6 may be necessary, including the installation of additional pumps where space is available, to mitigate beam-gas backgrounds in the ePIC detector. Preliminary detector background studies indicate that the resulting rates remain acceptable for the tracking system performance, though further evaluation is ongoing within the detector working group.

\subsection{Collimation Performance}
Following the approach used in LHC collimation studies, we compute the local cleaning inefficiency ($\eta$), defined as the fractional loss per unit length at a given longitudinal position ($S$) around the ring~\cite{redaelli2025beamcleaningcollimationsystems}:
\begin{equation}
    \eta(S) = \frac{N_\mathrm{loc}(S \rightarrow S + \Delta S)}{N_\mathrm{coll}} \frac{1}{\Delta S},
\end{equation}
where $N_\mathrm{loc}(S \rightarrow S + \Delta S)$ is the number of particles lost within the interval $\Delta S$, and $N_\mathrm{coll}$ is the number of particles intercepted by the primary collimators.

The last column of Table~\ref{tab:LossesLifetime} reports $\eta$ for the three ESR beam energies. For each energy, the local loss term $N_\mathrm{loc}$ is obtained by summing all IR6 losses listed in the fifth column, and $\Delta S = \SI{37.5}{m}$ corresponds to the length of the IR6 cryostat section. The denominator $N_\mathrm{coll}$ is taken as the total number of particles absorbed by the collimators, indicated by the blue markers in Fig.~\ref{fig:ring_beamlosses_rate}.

The resulting values quantify the fraction of the beam halo that is not absorbed by the collimation system and leaks to IR6. These inefficiencies will serve as key inputs for forthcoming quench-margin evaluations of the ESR SC magnets and for detailed detector background studies.

\section{\label{sec:FurtherStudiesOutlook}Further Studies and Outlook}
The current simulation model has several limitations, including the absence of crab cavities and detector solenoid fields in the lattice. Although preliminary estimates suggest that their impact on beam losses is negligible, their implementation is in progress to ensure completeness. The updated lattice designs will incorporate these missing elements and may require adjustments to magnet configurations to account for IR6 synchrotron radiation, HSR lattice modifications, or the installation of additional vacuum components.

Based on the current betatron phase-space coverage in IR4, which is nearly $2\pi$, we are confident that the new lattice designs will allow identification of suitable drift spaces for potential collimator relocation within the region shown in Fig.~\ref{fig:ir4_optics}, even in the presence of phase-advance differences between lattice versions as the ESR lattice design matures further.

Additional studies are planned to investigate machine failure scenarios and injection-related beam losses to assess whether additional collimators are necessary. Future lattice versions will also include machine error studies introducing realistic misalignments and magnetic field errors for more comprehensive evaluations.

The ePIC detector team is concurrently studying detector backgrounds to define acceptable beam loss limits from both radiation damage and performance degradation perspectives, including effects on electronics and tracking reconstruction.

Finally, the vacuum model used in the current simulations requires refinement. This includes a more accurate representation of gas composition, as well as the evolution of pressure profiles along the ring and over time, to improve the fidelity of future beam loss and lifetime estimates.

\section{\label{sec:Conclusions}Conclusions}
We have presented the first baseline design of the electron ring collimation system for the Electron–Ion Collider and demonstrated its effectiveness through high-statistics multi-turn tracking simulations. By integrating a dedicated betatron-collimation insertion into IR4 and optimizing its optics across all ESR beam energies, we have shown that a minimal system consisting of a single primary collimator per plane is sufficient to localize beam-gas and Touschek losses and suppress IR6 losses by one to two orders of magnitude, assuming operation after sufficient integrated beam dose has been accumulated to achieve low residual gas pressure. Importantly, this level of mitigation is achieved without compromising the machine acceptance or significantly affecting the multi-hour beam lifetime required for stable swap-out operation.

The results confirm that the proposed collimation insertion protects the SC final-focusing magnets and significantly reduces beam losses in the IR. The resulting loss levels are consistent with expectations for controlling detector background sources, although a full evaluation of detector backgrounds is beyond the scope of this work. The achieved heat-load densities remain below the preliminary cryogenic limits, establishing IR4 as a robust and sustainable location for beam-loss interception in the present lattice design.

This work also defines a clear roadmap for further refinements. Incorporation of the detector solenoid field, crab cavities, updated vacuum models, and realistic machine error distributions into the simulation model will complete the integration of the collimation system into the full ESR design. Combined with forthcoming radiation studies to evaluate cold mass loads and detector backgrounds, these efforts will lead to a fully validated, operations-ready collimation system for the ESR.

The design and results presented here now serve as the baseline for the ESR collimation system and have been adopted in ongoing lattice development and machine-protection studies. They provide a solid foundation for commissioning strategies and for future updates as the EIC moves toward construction and operation.

\section{\label{sec:Acknowledgments}Acknowledgments}
We thank our colleagues in the EIC Project and the ePIC Collaboration for many helpful discussions and for providing technical details on the experimental setup. We also express our sincere gratitude to the SuperKEKB-Belle~II team for their extensive efforts in understanding the dominant beam-loss processes in modern high-intensity electron-positron rings, which formed the basis for modeling the ESR beam-loss mechanisms. We gratefully acknowledge R.~Bruce, S.~Redaelli, and F.~Van~Der~Veken (CERN), members of the LHC Collimation and Xsuite developer teams, for their assistance and support. We extend special thanks to G.~Broggi (CERN) for his dedication and highly professional support with the Xsuite framework, and to J.~Unger (Cornell University) for cross-checking the Touschek lifetime calculations using the Bmad framework for earlier versions of the ESR lattice. We also thank the Jefferson Lab Scientific Computing Team for providing access to the JLab computing farm, which enabled the high-statistics, CPU-intensive simulations presented in this work.

This work was supported by the U.S. Department of Energy under contract number DE-SC0012704.

\bibliography{bibfile}

\end{document}